\documentclass[aps,pre,twocolumn,10pt,floatfix]{revtex4-1}

\usepackage{amsmath,bm,mathtools}
\usepackage{amssymb}
\usepackage{graphicx}
\usepackage[breaklinks]{hyperref}
\usepackage{xcolor}
\usepackage[utf8]{inputenc}
\usepackage{tikz}
\usepackage{braket}

\hypersetup{
    colorlinks,
    linkcolor={red!80!black},
    citecolor={green!70!black},
    urlcolor={blue!80!black}
}

\usepackage[]{natbib}

\PassOptionsToPackage{linktocpage}{hyperref}

\DeclareMathOperator{\Tr}{Tr}

\newcommand*{\diff}{\mathop{}\!\mathrm{d}}

\def\mean#1{\mathinner{\langle{#1}\rangle}}

\begin{document}

\title{Large deviations of the free energy in the \texorpdfstring{$p$}{p}-spin glass spherical model}
\date{\today}
\author{Mauro Pastore}
\email{mauro.pastore@unimi.it}
\author{Andrea Di Gioacchino}
\email{andrea.dgioacchino@gmail.com}
\affiliation{Dipartimento di Fisica, Universit\`a degli Studi di Milano}
\affiliation{INFN, Via Celoria 16, I-20133 Milan, Italy}
\author{Pietro Rotondo}
\email{pietrorotondo86@gmail.com}
\affiliation{School of Physics and Astronomy, University of Nottingham, Nottingham, NG7 2RD, UK}
\affiliation{Centre for the Mathematics and Theoretical Physics of Quantum Non-equilibrium Systems, University of Nottingham, Nottingham NG7 2RD, UK}

\begin{abstract}
We investigate the behavior of the rare fluctuations of the free energy in the $p$-spin spherical model, evaluating the corresponding rate function via the G\"artner-Ellis theorem. This approach requires the knowledge of the analytic continuation of the disorder-averaged replicated partition function to arbitrary real number of replicas.
In zero external magnetic field, we show via a one-step replica symmetry breaking (1RSB) calculation that the rate function is infinite for fluctuations of the free energy above its typical value, corresponding to an anomalous, super-extensive suppression of rare fluctuations. We extend this calculation to non-zero magnetic field, showing that in this case this \emph{very large} deviation disappears and we try to motivate this finding in light of a geometrical interpretation of the scaled cumulant generating function.
\end{abstract}


\maketitle

\section{Disordered systems and large deviations}
\label{sec:intro}

The theory of disordered systems has been mainly developed to describe the typical behavior of physical observables. However, as it has been argued since the early days of the subject, one can employ spin glass techniques in a more general setting, to estimate probability distributions \cite{TD_old} and fluctuations around the typical values \cite{Tanaka,Crisanti_1990} of quantities of interest. More recently, Rivoire \cite{RivoireCAV}, Parisi and Rizzo \cite{rizzoPRL,rizzoPRB2008,rizzoPRB2010,rizzoJPA} and others \cite{AndREM,nakhukSK,nakhukPRE} followed this line of thought, providing a bridge between spin glasses (and disordered systems more in general, as in \cite{MalatestaLDMatch}) 
and the theory of large deviations, that deals with rare events whose probability decays exponentially in the system size. This topic, which is the natural framework to set statistical mechanics in a mathematical perspective, has recently been the subject of a comprehensive and pedagogical review by Touchette \cite{touchLD}, as well as of intensive efforts in non-equilibrium statistical physics \cite{Carlos:Chaos:2019}.

The key quantity providing the bridge is of course very familiar to spin glass physicists and is given by:
\begin{equation}
G(k) = \lim_{N\to\infty} -\frac{1}{\beta N} \log \mean{Z_N^k}\,,
\label{eq:SCGF}
\end{equation}
where $Z_N$ is the partition function for a system of size $N$ and $\mean{\cdots}$ is the average over disorder. The argument of the logarithm is the averaged replicated partition function and $k$ is the so-called replica index. From the viewpoint of large deviation theory, $G(k)$ is simply related to the scaled cumulant generating function (SCGF) of the free energy $f=\lim_{N\to\infty}f_N$ by
\begin{equation}
\psi(k) = \lim_{N\to\infty} \frac{\log \mean{e^{k N f_N}}}{N} = -\beta G(-k/\beta) \,.
\label{eq:psifromG}
\end{equation}

If the SCGF is differentiable one can show that the probability $P (f)$ describing the fluctuations of the free energy satisfies a large deviation principle:
\begin{equation}\label{ldp}
P(f_N \in [x, x + \diff x]) \sim e^{-N I(x)} \diff x\,
\end{equation}
and the rate function $I(x)$ is given by the Legendre transform of the SCGF:
\begin{equation}
I (x) = \sup_{k \in \mathbb{R}} \left[ x k - \psi (k) \right]\,.
\label{eq:GE}
\end{equation}
This is an application of a standard result of large deviation theory known as G\"artner-Ellis theorem and describes the rare fluctuations of the free-energy around the typical value $f_{\textrm{typ}}$, which corresponds to the special point of the rate function $I (f_{\textrm{typ}}) = 0$.

From the disordered systems perspective, most of the standard results of spin glass theory obtained within the replica method concern only the very special limit $k \rightarrow 0$, since $f_{\textrm{typ}} = \mean{f} = \psi'(0)$, whereas to obtain the full form of $I (x)$ that describes arbitrary rare fluctuations of the free-energy one needs to work out the SCGF for finite replica index $k$. This problem is clearly equivalent to determine the full analytical continuation of the averaged replicated partition function from integer to real number of replicas $k$ and it was extensively investigated in the early stage of the research in disordered systems in order to understand the manifestation of the (at that time surprising) mechanism of replica symmetry breaking \cite{parisiRSB}. Since these results are particularly interesting from the more modern large deviation viewpoint, we briefly mention the main ones in the following.

Van Hemmen and Palmer~\cite{vanH-P} were the first ones to observe that the expression in Eq.~\eqref{eq:SCGF} must be a convex function of the replica index $k$, as can be proven by exploiting H\"older's inequality. Shortly later, Rammal~\cite{Rammal} added that $\psi(k)/k$ must be monotonic, which is actually a necessary condition for the convexity of $\psi(k)$. However the replica symmetric (RS) ansatz, which provides the most obvious analytical continuation to real $k$ of the replicated partition function, gives often a trial SCGF which is not convex, or such that $\psi(k)/k$ is not monotonic. This problem has been analyzed for the first time in the context of the Sherrington-Kirkpatrick model. After Parisi introduced his remarkable hierarchical scheme for replica symmetry breaking, Kondor \cite{kond} argued that his full RSB solution was very likely to provide a good analytical continuation of Eq.~\eqref{eq:SCGF}, not only around $k=0$.

These results may be considered nowadays as the initial stage of a work that attempted to give mathematical soundness to the replica method. Although this vaste program is mostly unfinished, Parisi and Rizzo realized that the original analysis presented by Kondor is fundamental to investigate the large deviations of the free-energy in the SK model. Large deviations have been examined only for a few other spin glass models: Gardner and Derrida discussed the form of the SCGF in the random energy model (REM) in a seminal paper \cite{gardnerREM}, and many rigorous results have been established later on \cite{fedrigo2007}; Ogure and Kabashima \cite{ogkabPTP,ogkabJSM1,ogkabJSM2} considered analyticity with respect to the replica number in more general REM-like models; Nakajima and Hukushima investigated the $p$-body SK model \cite{nakhukSK} and dilute finite-connectivity spin glasses \cite{nakhukPRE} to specifically address the form of the SCGF for models where one-step replica symmetry breaking (1RSB) is exact.  

In this manuscript we add one more concrete example to this list, considering the $p$-spin spherical model \cite{CS}. In zero external magnetic field, we show that the 1RSB calculation at finite $k$ produces a SCGF with a linear behavior below a certain value $k_c$; a nice geometrical interpretation of this, dating back to Kondor's work on the SK model \cite{kond}, is discussed. Accordingly, the rate function is infinite for fluctuations of the free energy above its typical value, which are then more than exponentially suppressed in $N$ with respect to the standard case described by Eq.~\eqref{ldp}. This property, which is commonly described stating that the free energy has a ``very-large'' deviation behavior for positive fluctuations, is present in several other spin glass problems, as discussed for example in \cite{rizzoPRB2010}, and, more generally, in other systems showing extreme value statistics \cite{AndREM}. In some of the early literature \cite{Dotsenko_1994}, this feature is also called ``overfrustration''.

The situation changes dramatically when a small external magnetic field is applied: the rate function is finite everywhere, although highly asymmetric around the typical value, and so the very-large deviation feature disappears. We explain intuitively the reason of this change of regime in light of the geometrical interpretation discussed for the case without magnetic field, and argue that the introduction of a magnetic field could act as procedure to regularize the anomalous scaling of the large deviation principle for this kind of systems.

The manuscript is organized as follows: in Section \ref{sec:nofield} we derive the SCGF for the $p$-spin spherical model without magnetic field; then we employ the G\"artner-Ellis theorem to compute the corresponding rate function in the high- and low-temperature phases. In Section \ref{sec:field} we generalize the results to non-zero magnetic field and compare the SCGF and the rate function obtained with those of the previous case. In Section \ref{sec:conclusions} we summarize our results and discuss possible future directions.
Finally, in Appendix \ref{app:rammal} we discuss the details of the geometrical interpretation of the 1RSB ansatz.

\section{Large deviations of the \texorpdfstring{$p$}{p}-spin spherical model free energy}
\label{sec:nofield}

\begin{figure}
\centering
\includegraphics[scale=1.07]{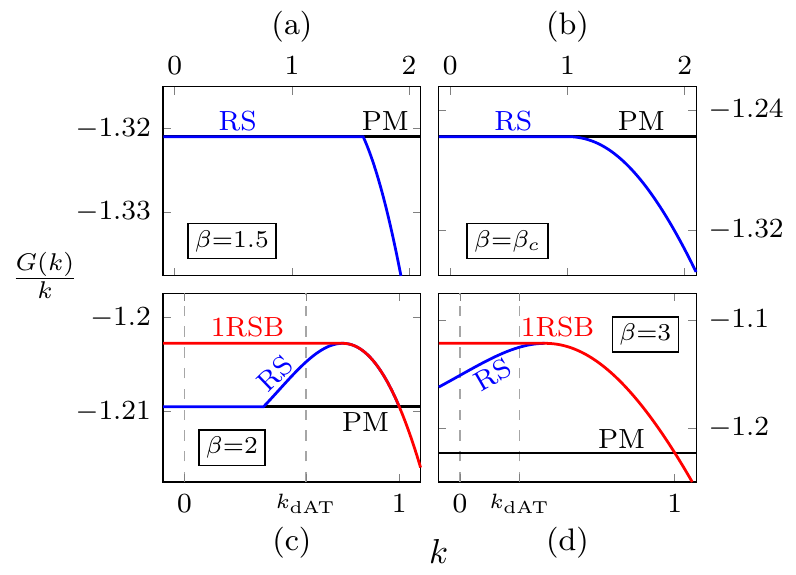}
\caption{The function $G(k)/k$ for the ($p=3$)-spin in zero external magnetic field, for different values of $\beta$. (a) At high temperature ($\beta=1.5$) the 1-RSB anstatz coincides with the RS one (blue curve); the solution joins the paramagnetic line (in black) in a point $k_c>1$, where the function is not differentiable. (b) At $\beta=\beta_c\approx 1.706$, the junction is in $k_c=1$ and becomes smooth. For $\beta=2$ (c) and $\beta=3$ (d), the 1RSB solution (red curve) departs from the RS one and becomes a straight line for all the $k<k_c$, which is the point where the RS function loses its monotonicity. The critical value $k_c$ approaches zero for $\beta\to\infty$.}
\label{fig:G(k)/k}
\end{figure}

The $p$-spin glass spherical model consists of a $p$-body interaction of $N$ continuous spins with the following Hamiltonian:
\begin{equation}
H_p =- \sum_{1\leq i_1 <  i_2 < \cdots < i_p \leq N} J_{i_1\cdots i_p} \sigma_{i_1}\sigma_{i_2}\cdots \sigma_{i_p}\,,
\label{eq:H}
\end{equation} 
where the $J$-couplings are independent quenched random variables normally distributed with zero mean and variance 
\begin{equation}
\braket{J^2_{i_1 \cdots  i_p}} = \frac{J^2 p!}{2 N^{p-1}}\,,
\end{equation} 
while the spins are real variables with range in $(-\infty,\infty)$ subject to a global spherical constraint such that the measure is
\begin{equation}
\Tr_{\sigma} \equiv 2 \sqrt{N} \int_{-\infty}^{\infty} \prod_{i=1}^N d\sigma_i \,\delta \left( \sum_{i=1}^N\sigma_i^2 - N\right)\,.
\label{eq:trace_spherical}
\end{equation}
These scalings guarantee the extensivity of the free energy. Its exact typical value is obtained within the replica formalism and a 1RSB ansatz, as done in the seminal work \cite{CS} by Crisanti and Sommers (CS). 

In the following we will analyze the large deviations of the free energy of this model. For sake of brevity, we will not reproduce all the steps of CS, whose analysis we will extend here to any real finite number of replicas.

\subsection{From replicas to the scaled cumulant generating function}

We start our analysis from Eq.~(3.16) of \cite{CS} with null magnetic field. Accordingly, the partition function is (up to finite-size corrections in $N$):
\begin{equation}\label{eq:Zk_0}
\mean{Z_N^k} = \int_{\mathbf{q}>0} \prod_{\alpha < \beta} \diff q_{\alpha \beta} \, e^{-N g(\mathbf{q})}\,,
\end{equation}
where 
\begin{equation}
g(\mathbf q) = -\frac{(\beta J)^2}{4} \sum_{\alpha, \beta = 1}^k q_{\alpha \beta}^p - \frac{1}{2}\log \det \mathbf{q} - k s(\infty)\,.
\label{eq:g0}
\end{equation}
and $s(\infty) = [1+\log (2\pi)]/2$ is the entropy density in the infinite temperature limit.
To evaluate the integrals on $q_{\alpha \beta}$ we use the saddle point method together with the 1RSB ansatz, which is formulated in terms of the three parameters $(q_1, q_0, m)$:
\begin{equation}
q_{\alpha\beta} = (1 - q_1)\delta_{\alpha\beta} + (q_1 - q_0)\epsilon_{\alpha\beta} + q_0
\label{eq:q1RSB_compoments}
\end{equation}
with $\epsilon_{\alpha\beta}$ defined as
\begin{equation}
\epsilon_{\alpha\beta} = \begin{cases}
1 &\text{if $\alpha$, $\beta$ are in a diagonal $m \times m$ block,}\\
0 & \text{otherwise.}
\end{cases}
\end{equation}
The eigenvalues of $\mathbf{q}$, with the respective degeneracies, are
\begin{equation}
\begin{aligned}
\eta_0 &= 1-q_1 &				& \text{deg.} = k(m-1)/m \\
\eta_1 &= 1-(1-m)q_1-m q_0 &	& \text{deg.} =  k/m-1\\
\eta_2 &= 1-(1-m)q_1 -(m-k)q_0  & & \text{deg.} = 1
\end{aligned}
\end{equation} 
Using this and inserting the ansatz \eqref{eq:q1RSB_compoments} in \eqref{eq:g0} we find
\begin{equation}\label{eq:g_h_0}
\begin{aligned}
g(k;q_0,q_1,m) = &- \frac{(\beta J)^2}{4}k\left[1 + (m-1)q_1^p + (k-m)q_0^p \right]\\
&-\frac{k(m-1)}{2m}\log\left({\eta_0}\right) - \frac{k}{2m}\log\left({\eta_1}\right)\\
&- \frac{1}{2}\log\left(1+ \frac{ k q_0}{\eta_1}\right)- k s(\infty)  \,.
\end{aligned}
\end{equation}
This functional is evaluated numerically at the saddle point $(\bar{q}_1, \bar{q}_0, \bar{m})$ for the 1RSB parameters for each value of $k$. The three parameters take values in the domains $q_1\in[0,1]$, $q_0\in[0,q_1]$, $m\in[1,k]$ (if $k>1$) or $m\in[k,1]$ (otherwise), and for $k<1$ the saddle point is obtained with a maximization of the functional instead of a minimization, as usual in replica theory. Using Eq.~\eqref{eq:psifromG}, we obtain a SCGF $\psi(k)$ which becomes linear above a certain value $k=k_c$, depending on temperature. 
To ease the visualization of this feature, in Fig.~\ref{fig:G(k)/k} we plot the function
$G(k)/k = g(k;\bar{q}_1, \bar{q}_0, \bar{m})/(k\beta)$ which, when $\psi(k)$ is linear, intersects the vertical axis in $f_{\text{typ}}$. The figure does not change qualitatively for $p\ge 3$.

The $p=2$ case at low temperature is different: the 1RSB ansatz reduces to the RS one (that is, $\bar{q}_1=\bar{q}_0$) as long as $k \geq 0$, therefore the typical values of all the thermodynamic quantities are obtained under the RS ansatz \cite{PhysRevLett.36.1217}. On the opposite, for $k<0$ we need to introduce again the 1RSB ansatz which, as in the $p \geq 3$ case, gives the linear behavior of the SCGF. In other words, $k_c = 0$ for the 2-spin spherical model for all $\beta>\beta_c$.

Before turning to the evaluation of the rate function, we discuss an interesting geometrical interpretation of the SCGF shape. To this aim, let us consider the RS ansatz (that is, Eq.~\eqref{eq:g_h_0} with $q_1=q_0=q$ and $m=1$).
As we can see in Fig.~\ref{fig:G(k)/k}, the RS solution (blue curve) is non-monotonic for $\beta>\beta_c$.
On the other hand, one can prove that $G(k)/k$ has to be a monotonic quantity, therefore the RS solution can be ruled out. 
We can check that the 1RSB solution gives a perfectly fine monotonic $G(k)/k$ (red curve in Fig.~\ref{fig:G(k)/k}), as one could expect due to the fact that this ansatz gives the correct typical free energy for this model. Interestingly, however, exactly the same monotonic curve can be obtained by using a much simpler geometric construction: just consider the RS solution, which is the right one for large $k$, and when $G(k)/k$  starts to be non-monotonic continue with a straight horizontal line (in the $G(k)/k$ vs $k$ plot). 
This construction actually dates back to Rammal \cite{Rammal} and is discussed in more detail in Appendix~\ref{app:rammal}. 
Here we limit ourselves to notice that $G(k)/k$ obtained by using the 1RSB ansatz or the Rammal construction are the same because of the following facts: (i) for $k > k_c$ the 1RSB and RS ansatz\"e coincide ($\bar{q}_1 = \bar{q}_0 = q \neq 0$) and $k_c$ is exactly the point where $G(k)/k$ is not monotonic anymore if one uses the RS ansatz; (ii) from the saddle point equations obtained by extremizing Eq.~\eqref{eq:g_h_0} when $k < k_c$, one obtains $\bar{q}_0=0$; (iii) the remaining saddle point equations fix $q_1$ and $m$, and one can see that these equations are identical to those needed to perform the Rammal construction, which fix the point $k_c$ and the parameter of the RS ansatz $q$.

\subsection{Rate function and very large deviations}

\begin{figure}
\centering
\includegraphics[scale=1.19]{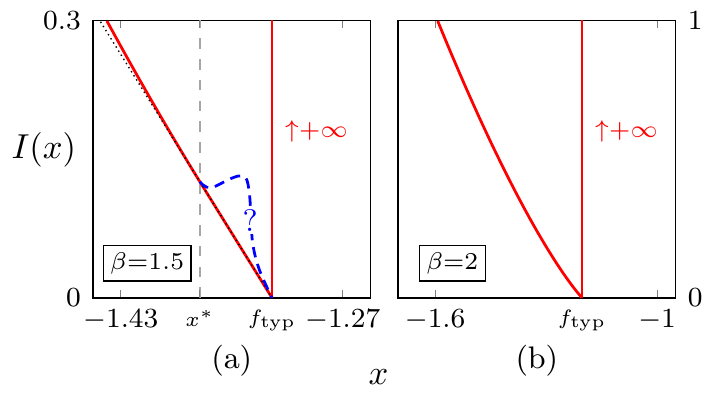}
\caption{Rate function of the free energy for the ($p=3$)-spin in zero external magnetic field, for different values of $\beta$. The fluctuations above the typical value correspond to the linear part of the SCGF, so that the Legendre transformation gives an infinite rate function. The fluctuations below the typical value are described by the branch in red. For $\beta=1.5<\beta_c$ (a), as the SCGF is not differentiable, 
we obtain only the convex-hull of the true rate function; in the interval $[x^*,f_{\text{typ}}]$, where our result gives a straight segment (the part of the curve overlapping the dotted line), the true, unknown rate function is represented by the curve in blue. For $\beta=2>\beta_c$ (b) the SCGF is smooth and the G\"artner-Ellis theorem applies.}
\label{fig:rate}
\end{figure}

Starting from the SCGF evaluated in the last section, we perform a numerical Legendre transformation to obtain the rate function according to Eq.~\eqref{eq:GE}. The result is shown in Fig.~\ref{fig:rate} for different values of $\beta$. The rate function displays the following behavior: 
\begin{itemize}
\item for $x=f_{\text{typ}}$, it is null as expected; 
\item for $x<f_{\text{typ}}$, $I(x)$ is finite, indicating that a regular large deviation principle holds for fluctuations below the typical value. When $\beta>\beta_c$ the SCGF is smooth, so we obtain the rate function via the Gartner-Ellis theorem. On the other hand, when $\beta<\beta_c$ the SCGF is not differentiable in a point (see Fig.~\ref{fig:G(k)/k}), so we are only able to obtain the convex hull of the rate function (see Fig.~\ref{fig:rate});
\item for $x>f_{\text{typ}}$, $I(x)=+\infty$. This is due to the linear behavior of the SCGF below $k_c$ discussed in the previous section and it is a signature of an anomalous scaling with $N$ of the rare fluctuations above the typical value.
\end{itemize}
An ambitious goal would be the identification of the correct behavior with $N$ of these very large deviations. Indeed, a more general way of stating a large deviation principle is
\begin{equation}\label{ldp2}
P(f_N \in [x, x + \diff x]) \sim 
\begin{cases}
e^{- a_N I_-(x)}\diff x &	\text{if $x\le f_{\text{typ}}$}\,,\\
e^{- b_N I_+(x)}\diff x	&	\text{if $x>f_{\text{typ}}$}\,,\\
\end{cases}
\end{equation}
where $a_N, b_N \to \infty$ when $N\to\infty$. In other words, the fluctuations resulting in values of $x$ lower than $f_{\text{typ}}$ are given by the rate function $I_-(x)$, while those resulting in values larger than $f_{\text{typ}}$ have rate function $I_+(x)$, but with different scalings $a_N$, $b_N$. In our case, we have $a_N \sim N$, then the rate function defined in Eq.~\eqref{ldp} can be written as
\begin{equation}
I(x) \sim \begin{cases}
I_-(x) &	\text{if $x\le f_{\text{typ}}$}\,,\\
\frac{b_N}{N} I_+(x)	&	\text{if $x>f_{\text{typ}}$}\,,\\
\end{cases}
\end{equation}
with  $b_N/N \to \infty$. For this reason, fluctuations above the typical value are referred to as ``very large deviations''. The physical explanation of the substantial difference in scaling of the deviations of thermodynamical quantities below and above their typical values resides in the different number of elementary degrees of freedom involved to obtain the corresponding fluctuation: while in the first case it is sufficient that only one of the elementary variables assumes an anomalous value below its typical, the others being fixed, in the second case all the variables have to fluctuate, a joint event with probability heavily suppressed with respect to the first one.

This argument shows the importance of the resolution of the anomalous scaling behavior leading to the very large deviations we explained above. In general, however, although the G\"artner-Ellis theorem can be extended to find rate functions for large deviation principles with arbitrary speed $a_N$, $b_N$, we lack techniques to compute the asymptotic scaling of $a_N$ and $b_N$ for large $N$, because of additional inputs needed to calculate the corresponding SCGF with a saddle-point approximation (for some other systems this problem has been solved with ad-hoc methods \cite{AndREM,DeanRM}, while in \cite{rizzoPRB2010} a method is proposed in the context of the SK model).

In the next section we present the main result of our work, which could be useful to study this anomalous kind of fluctuations also in other problems: through an extension of the replica calculation to the case with an external magnetic field, we are able to numerically check that the very large deviation effect disappears. More in detail, we obtain that with a magnetic field, no matter how small, not only $a_N \sim N$ as before, but also $b_N \sim N$.

\section{Large deviations of the \texorpdfstring{$p$}{p}-spin model in a magnetic field}
\label{sec:field}

In this section we generalize the previous discussion to the case of non-zero magnetic field. The Hamiltonian for the model is
\begin{equation}
H = H_p  - h\sum_{i = 1}^N \sigma_i \, ,
\end{equation}
where $H_p$ is given in \eqref{eq:H} and $h$ represents an external magnetic field coupled with the spins. 

The computation of the SCGF at $h\neq 0$ goes beyond the approach of the work by Crisanti and Sommers, who only considered the typical case. In contrast to the problem with $h=0$, where the finite-$k$ calculation consists of a quite straightforward generalization of the standard one, here a more substantial effort is needed to extend the $k=0$ result. 

The starting point is Eq.~(3.8) of \cite{CS}, which we report here for convenience:
\begin{equation}
\begin{aligned}
\mean{Z^k_N} = &\int_{\mathbf{q}>0} \prod_{\alpha<\beta} \diff q_{\alpha\beta} \int_{-i \infty}^{+i \infty} \prod_{\alpha<\beta} \frac{N}{2\pi i} \diff \lambda_{\alpha\beta}\\
&\cdot \int_{-i \infty}^{+i \infty} \prod_{\alpha} \frac{\sqrt{N}}{2\pi i} \diff \lambda_{\alpha\alpha} \, e^{-N g(\mathbf{q},\bm{\lambda})}\,,
\end{aligned}
\end{equation}
where the entries of the $\bm{\lambda}$ matrix are auxiliary variables enforcing the constraints defining the overlap matrix
\begin{equation}
q_{\alpha\beta} = \frac{1}{N} \sum_{i=1}^N \sigma_{i\alpha}\sigma_{i\beta}
\end{equation}
and the spherical constraint in \eqref{eq:trace_spherical}. In the presence of a magnetic field, the saddle-point integration in the $\lambda$-variables is not straightforward as to obtain \eqref{eq:g_h_0}. The full expression of $g(\mathbf{q},\boldsymbol{\lambda})$ before the $\mathbf{\lambda}$-integration reads:
\begin{multline}
g(\mathbf{q},\boldsymbol{\lambda}) = -\frac{(\beta J)^2}{4}\sum_{\alpha, \beta=1}^k q_{\alpha \beta}^p +\frac{1}{2}\sum_{\alpha, \beta = 1}^k \lambda_{\alpha \beta} q_{\alpha \beta}\\
+\frac{1}{2}\log \det \left(-\bm{\lambda}\right)
+\frac{(\beta h)^2}{2}\sum_{\alpha, \beta=1}^k \left(\bm{\lambda}^{-1}\right)_{\alpha \beta} - \frac{k}{2} \log(2\pi) \,.
\label{eq:G_1RSB_full}
\end{multline}
Derivation with respect to $\lambda_{\alpha\beta}$ leads to the following saddle-point equations:
\begin{equation}
q_{\alpha \beta} + \left(\bm{\lambda}^{-1}\right)_{\alpha \beta} -(\beta h)^2 \sum_{\gamma,\delta=1}^k \left(\bm{\lambda}^{-1}\right)_{\gamma \alpha} \left(\bm{\lambda}^{-1}\right)_{\beta \delta} = 0\,,
\label{eq:splambda}
\end{equation}
where we have used the identity:
\begin{equation}
\frac{\partial \left(\bm{\lambda}^{-1}\right)_{\gamma \delta}}{\partial\lambda_{\alpha \beta}} =- \left(\bm{\lambda}^{-1}\right)_{\gamma \alpha} \left(\bm{\lambda}^{-1}\right)_{\beta \delta}\,.
\end{equation}
Equations \eqref{eq:splambda} are solved via successive contractions of the replica indices: a double summation over $\alpha$, $\beta$ leads to an equation for the scalar $\sum_{\alpha \beta}\left(\bm{\lambda}^{-1}\right)_{\alpha \beta}$ with solutions:
\begin{equation}
\begin{aligned}
\sum_{\alpha, \beta=1}^k\left(\bm{\lambda}^{-1}\right)_{\alpha \beta} &= \frac{1\pm\sqrt{1+4(\beta h)^2 q_s}}{2(\beta h)^2} \equiv l_{\pm}\,, \\
q_s &=\sum_{\alpha \beta}q_{\alpha \beta}\,.
\end{aligned}
\end{equation}
Similarly, a single contraction gives:
\begin{equation}
\sum_{\alpha}\left(\bm{\lambda}^{-1}\right)_{\alpha \beta} = -\frac{\sum_{\alpha} q_{\alpha \beta}}{1-(\beta h)^2 l_{\pm}}\,,
\end{equation}
and finally 
\begin{equation}
\left(\bm{\lambda}^{-1}\right)_{\alpha \beta}  = -q_{\alpha \beta} +\frac{(\beta h)^2 \sum_{\gamma}q_{\gamma \alpha}\sum_{\delta} q_{\delta \beta}}{\left[1-(\beta h)^2 l_{\pm}\right]^2}\,.
\label{eq:splambda_sol}
\end{equation}
Given the 1RSB ansatz \eqref{eq:q1RSB_compoments}, $q_{\alpha\beta}$ has $k$ elements 1 on the diagonal, $m (m-1) k/m$ elements $q_1$ in the internal blocks, the remaining $k^2 - k - k(m-1)$ elements $q_0$, so
\begin{equation}
q_s = k + k (m - 1) q_1 + k(k - m) q_0 = k \eta_2
\end{equation}
Every row (column) contains the same elements, so
\begin{equation}
q_r \equiv \sum_\beta q_{\alpha\beta} = 1 + (m - 1) q_1 + (k - m) q_0=\eta_2  \qquad \forall\, \alpha\,.
\end{equation}
To find which of the parameters $l_\pm$ in Eq.~\eqref{eq:splambda_sol} is the right one, we can perform the limit $k\to 0$:
\begin{equation}
\begin{aligned}
q_s &\to 0\,,\qquad  q_r \to 1 + (m - 1) q_1 - m q_0 \,, \\
l_\pm(q_s) &\to l_\pm(0) = \begin{cases}
1/(\beta h)^2 \,,\\
0\,,
\end{cases}
\end{aligned}
\end{equation}
so that $\bm{\lambda}$ has a finite limit only with $l_-$, for which the saddle-point equations become
\begin{equation}
\left(\bm{\lambda}^{-1}\right)_{\alpha \beta}  = -q_{\alpha \beta} +\hat{q}_-\,,
\end{equation}
where
\begin{align}
\hat{q}_-  &= \frac{4(\beta h)^2\eta_2^2}{\left[ 1 + \sqrt{1 + 4(\beta h)^2 k \eta_2}\right]^2}  \label{eq:hatq} \\
&\underset{k\to 0}{\longrightarrow}  (\beta h)^2 \left[ 1 + (m - 1) q_1 - m q_0\right]^2 =  (\beta h)^2 \eta_1^2 \,.
\end{align}
The structure is the same as the one of $q_{\alpha\beta}$, with a constant added to each entry. Thus, the entries of $\bm{\lambda}^{-1}$
can be written as
\begin{equation}
\left(\bm{\lambda}^{-1}\right)_{\alpha \beta} = (q_1 - 1)\delta_{\alpha\beta} + (q_0 - q_1)\epsilon_{\alpha\beta} - q_0 + \hat{q}_{-}\,.
\end{equation}
It is also easy to see, inverting a matrix with a 1RSB structure, that
\begin{equation}
\lambda_{\alpha\beta} = -\frac{1}{\eta_0} \delta_{\alpha\beta} + \frac{q_1 - q_0}{\eta_0 \eta_1}\epsilon_{\alpha\beta} + \frac{q_0 - \hat{q}_-}{\eta_1\left(\eta_2 - k \hat{q}_-\right)} 
\end{equation}
and that $\bm{\lambda}$ has eigenvalues
\begin{equation}
\begin{aligned}
\kappa_0 &= -1/\eta_0  &  & \text{deg.} = n(m-1)/m \,,\\
\kappa_1 &= -1/\eta_1 & & \text{deg.} = n/m-1 \,,\\
\kappa_2 &= -1/\left(\eta_2 - k \hat{q}_- \right) \qquad & & \text{deg.} = 1 \,.
\end{aligned}
\end{equation}
The next step is to evaluate the trace appearing in \eqref{eq:G_1RSB_full}:
\begin{equation}
\Tr \left(\bm{\lambda} \times \mathbf{q} \right) = - k \left(1 + \frac{\hat{q}_-}{ \eta_2 -  n\hat{q}_{-}} \right)\, .
\end{equation}
Using all these ingredients, we can write the functional $g(\mathbf{q})$ in the 1RSB ansatz for finite $k$:
\begin{equation}\label{eq:g_h_non_0}
\begin{split}
g(k;q_0,q_1,m) = &-\frac{(\beta J)^2}{4}k\left[1 + (m-1)q_1^p + (k-m)q_0^p \right]\\
& -  \frac{k\hat{q}_-}{2( \eta_2 -  k\hat{q}_{-})} -\frac{k(m-1)}{2m}\log\left({\eta_0}\right)\\
&-\frac{k}{2m}\log\left({\eta_1}\right) - \frac{1}{2}\log\!\left(1 + \frac{k(q_0- \hat{q}_-)}{\eta_1}\right)\\
& -\frac{(\beta h)^2}{2} k \left(\eta_2 - k\hat{q}_-  \right) -k s(+\infty)\,.
\end{split}
\end{equation} 

\begin{figure}
\centering
\includegraphics[scale=1.06]{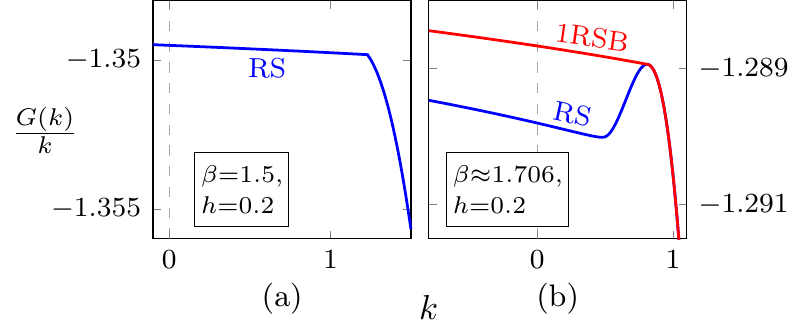}
\caption{The function $G(k)/k$ for the ($p=3$)-spin in a magnetic field $h=0.2$, for different values of $\beta$: (a) $\beta = 1.5<\beta_c(h)$, (b) $\beta=\beta_c(h=0)>\beta_c(h)$. The application of a magnetic field washes out the linear behavior at small $k$ observed in zero magnetic field.
}
\label{fig:scgf_h_non_0}
\end{figure}

As in the previous section, we numerically obtain and plot, in Fig.~\ref{fig:scgf_h_non_0}, $G(k)/k = g(k;\bar{q}_1, \bar{q}_0, \bar{m})/(k\beta)$, where again $\bar{q}_1, \bar{q}_0, \bar{m}$ are the solutions of the saddle point equations, obtained by extremization of Eq.~\eqref{eq:g_h_non_0}. The most striking feature of these plots is the difference from those represented in Fig.~\ref{fig:G(k)/k}: the linear behavior is replaced by curves (again given by the 1RSB ansatz) with non-null derivative.
Let us analyze more closely what is going on and why the external magnetic field is modifying the behavior of the system.
As discussed in the last part of Sec.~\ref{sec:nofield}, one can apply the Rammal construction to correct the non-monotonic behavior of the RS version of $G(k)/k$ (plotted as a blue curve in Fig.~\ref{fig:scgf_h_non_0}). Exactly as in the $h=0$ case, the resulting function will be monotonic and linear, which is the smooth continuation of $G(k)/k$ from $k_m$, the point where it loses its monotonicity. However, as one can see from Fig.~\ref{fig:scgf_h_non_0}), the result will not be the 1RSB solution.
This difference from the $h=0$ case can be seen as a consequence of the saddle point equations: now the equation for $q_0$ is non-trivial and so either $\bar{q}_0, \bar{q}_1$ and $\bar{m}$ depends on $k$ also in the 1RSB phase, giving rise to the non-constant behavior of $G(k)/k$ also for $k<k_c$. 
It is worth mentioning another point: when $h=0$, the critical point $k_c$ where the 1RSB solution departs from the RS one, coincides with $k_m$, the point where $G(k)/k$ obtained by the RS ansatz loses its monotonicity. Differently, with $h\neq 0$, we have that $k_c>k_m$ for $\beta>\beta_c$, so that the 1RSB branch departs from the RS one above $k_m$.
Finally, we numerically checked that the shape of $G(k)/k$ below $k_c$ depends on $p$.

\begin{figure}
\centering
\includegraphics[scale=1.06]{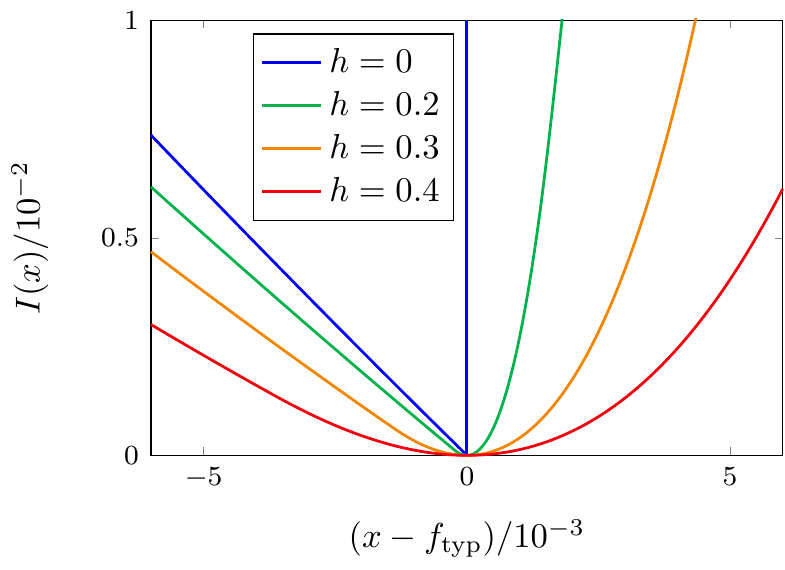}
\caption{Rate function of the free energy for the ($p=3$)-spin at $\beta = 3$, for different values of the external magnetic field. The infinite branch of the rate functions in Fig.~\ref{fig:rate} is replaced by a curve gradually less steep as the magnetic field is increased.}\label{fig:rate_h_non_0}
\end{figure}

This change in the SCGF has an important effect, in turn, on the rate function: performing the numerical Legendre transformation of the SCGF we now obtain a continuous curve, meaning that very rare fluctuations are washed out, see Fig.~\ref{fig:rate_h_non_0}. In other words, now the two quantities $a_N$ and $b_N$ introduced in Eq.~\eqref{ldp2} are such that $a_N \sim N$ and $b_N \sim N$. This effect is present also for very small magnetic field, even though the rate function is more and more asymmetrical around $x=f_{\text{typ}}$ as we decrease $h$.

\section{Discussion}\label{sec:conclusions}
In this manuscript we analyzed the behavior of the large (and very large) deviations of the free energy for the spherical $p$-spin model, exploiting the G\"artner-Ellis theorem to obtain the rate function. 
Without external magnetic field, we are able to compute the rate function in the spin-glass phase, while in the paramagnetic phase we obtain its convex hull, due to the non-differentiability of the SCGF. 
As a result, we have a standard large deviation principle for fluctuations below the typical value of the free energy, that is they are depressed exponentially in the size of the system. On the other hand, fluctuations above the typical value have a different behavior, being suppressed more than exponentially, and the corresponding rate function is infinite. 
When a magnetic field is applied, this anomalous very large deviation disappears and the rate function is finite everywhere. Since this remains true even if the field is very small, an open question is whether this effect can be exploited to obtain insights on the very large fluctuations, by sending the magnetic field to zero carefully choosing its dependence on the system size. 

In addition, we provided a geometrical interpretation to support our numerical findings. Indeed we showed, as noticed previously in the literature for different models, that for $h=0$ the Rammal construction is equivalent to the 1RSB ansatz. However, we also showed that this is due to the simple structure of the 1RSB ansatz without external magnetic field, where one can immediately fix one of the 1RSB parameters. When a magnetic field is applied, all the parameters have non-trivial values (which we obtained numerically by solving the saddle point equations) and the Rammal construction, which gives in turn the infinite-rate-function behavior, fails.
Another interesting question is whether it is possible to generalize the geometrical construction by Rammal to correct in the right way the RS solution not only for $h=0$, but also when $h\neq 0$.

\begin{acknowledgments}
The authors would like to thank Enrico Malatesta and Sergio Caracciolo for the useful discussions and suggestions.
\end{acknowledgments}

\appendix
\section{Rammal construction}
\label{app:rammal}

In this appendix we report the details of the geometrical construction reproducing the solution for the SCGF obtained with a 1RSB ansatz with $q_0 = 0$. The following observations are traced back to Rammal's work \cite{Rammal} and can be found in \cite{kond} (similar considerations in \cite{ogkabPTP,nakhukSK,nakhukPRE}). We reproduce here the reasoning not only as an historical curiosity: first of all, we see it as an enlightening approach to the problem of the continuation of the replicated partition function to real number of replicas, particularly suitable for a finite $k$ analysis. Moreover, we note that this interpretation, whenever it works, gives a flavor of ``uniqueness'' (though not in a strict mathematical sense) to the resulting solution, being based only on the properties of convexity and extremality that the function $\psi(k)$ must have. In this respect, a generalization of this result would be of great interest in order to better understand the necessity of Parisi hierarchical RSB procedure, which has been dubbed as ``magic'' even in relatively recent works, like \cite{Dotsenko_2011}; however, a true geometrical interpretation of the full machinery of RSB, beyond the simple case considered here, still lacks. Finally, in the context of this paper we are able to show a case where the construction gives the correct answer (the $p$-spin spherical model at zero external magnetic field) and a case where it fails (when the field is switched on).

Some important properties of the function \eqref{eq:psifromG} can be derived in full generality using its definition only. Applying the H\"older inequality to the probability measure over the disorder
\begin{equation}
\mean{XY} \le \mean{X^{1/k_1}}^{k_1} \mean{Y^{1/k_2}}^{k_2}\,,\quad 
\left\{\begin{aligned}
&0\le k_1,k_2\le 1\,,\\
&k_1 + k_2 = 1\,,
\end{aligned}\right.
\label{eq:LDP_Holder}
\end{equation}
with $X$, $Y$ some observables, it is easy to prove that $\psi(k)$ must be a convex function of $k$ (using $X=e^{\alpha k_1 N f_N}$, $Y=e^{(1-\alpha) k_2 N f_N}$ in the formula above, then taking the log and the large $N$ limit), and that $\psi(k)/k$ must be monotonic (using now $X=e^{k k_1 N A_N}$, $Y=1$).

Given that, the explicit evaluation is performed for each system within replica theory: an ansatz is imposed on the form of the replica overlap matrix, the number of replicas $k$ is then continued from integer to real values, the corresponding $G(k)$ is evaluated with the saddle-point method for large $N$ and finally a check is performed \textit{a posteriori} to verify its validity. In the SK model, the system originally considered by Rammal, at low temperatures the replica symmetric ansatz, which still gives the correct values of the positive integer momenta of the partition function, fails to produce a sensible solution for the SCGF at $k<1$, in at least three way:
\begin{itemize}
\item it becomes unstable under variations around the saddle point (de Almeida-Thouless instability~\cite{dAT}) below $k=k_{\text{dAT}}$;
\item it produces a $G(k)$ that is non-concave (and so a non-convex $\psi(k)$) around $k=k_{\text{conv}}$, meaning that $G''(k)$ changes sign at $k_{\text{conv}}$;
\item it produces a $G(k)/k$ that loses monotonicity a $k=k_m$.
\end{itemize}
In the SK model $k_{\text{dAT}}$ is the largest ($k_{\text{dAT}}>k_m>k_{\text{conv}}$), and so it is the first problem one encounters in extrapolating the RS solution from integer values of $k$. However, from the point of view of convexity and monotonicity alone, Rammal proposed to build a marginally monotone $G(k)/k$ in a minimal way, starting from the RS and simply keeping it constant below $k_m$ at the value $G(k_m)/k_m$. While the resulting function is not the correct one for the SK model, which needs a full RSB analysis to be solved, surprisingly enough for the spherical $p$-spin in zero magnetic field this approach reproduces the solution obtained with a 1RSB ansatz with $q_0 = 0$ (see Fig.~\ref{fig:G(k)/k}). Notice that in the present model the RS solution suffers from the same inconsistencies as in the SK model, but now $k_m$ is the largest of the three problematic points.

To convince the reader that the two approaches are actually equivalent we prove, as final part of this appendix, that without an external magnetic field the 1RSB solution of the spherical $p$-spin and the 
Rammal construction coincide. 
In order to obtain this result, we have to prove that:
\begin{itemize}
\item the 1RSB solution for $G(k)/k$ becomes a constant below $k=k_c$, which is defined as the point where the RS and 1RSB ans\"atze branch out, as we did in the main text;
\item this constant is the same as the one in the Rammal construction, that is $G(k_m)/k_m$;
\item the points $k_c$ and $k_m$ are the same.
\end{itemize}
As $k_c$ is the point where the RS solution is not optimal anymore, for $k<k_c$ we have $\bar{q}_0=0$, as discussed in \cite{CS}. Let us now consider Eq.~\eqref{eq:g_h_0} with $q_0=0$: differentiating with respect to $q_1$ and $m$ and setting the results equal to 0 we get the equations for $\bar{q}_1$ and $\bar{m}$, which read

\begin{equation}\label{eq:sp1}
\left\{
\begin{aligned}
    & \mu \, \bar{q}_1^{p-2} -  \frac{1}{(1-\bar{q}_1)(1-(1-\bar{m} )\bar{q}_1)} = 0\\
    & \frac{\mu}{2} \bar{m}^2 \bar{q}_1^p - \frac{1}{2} \log \left( 1+\frac{\bar{m}\,  \bar{q}_1}{1 - \bar{q}_1} \right)  + \frac{\bar{m}}{2} \frac{\bar{q}_1}{1-(1-\bar{m})\bar{q}_1} = 0
\end{aligned}
\right.
\end{equation}
where $\mu = p (\beta J)^2/2$. These equations can be solved numerically (as we did to obtain the plots in the main text), but to show our point here we do not really need the explicit solution. Indeed it is enough to notice that $\bar{m}$ and $\bar{q}_1$ do not depend on $k$ and therefore $g(k; 0, \bar{q}_1, \bar{m})/k$ is a constant.
Then, we need to check that it is the same constant as the one obtained by Rammal. Again starting from Eq.~\eqref{eq:g_h_0}, by putting $q_1=q_0=q$ we obtain the RS solution, which is
\begin{equation}
\begin{split}
g_{0} (k;q) = &- \frac{(\beta J )^2}{4} \left[k + k (k-1) q^p\right] - \frac{k-1}{2} \log(1-q)\\
    &- \frac{1}{2} \log\left[ 1 - (1-k) q \right] - k s(\infty).
\end{split}
\end{equation}
In this case, extremizing with respect to $q$, we have an equation which gives the RS solution on the saddle point, $\bar{q}$. To find $k_m$, we then require $\frac{\partial}{\partial k} g_{0} /k  = 0$. The two resulting equations are:
\begin{equation}\label{eq:sp2}
\left\{
\begin{aligned}
&     \mu \, \bar{q}^{p-2} -  \frac{1}{(1-\bar{q})\left[1-(1-k_m )\bar{q}\right]} = 0\\
& \frac{\mu}{2} k_m^2 \bar{q}^p - \frac{1}{2} \log \left( 1+\frac{k_m\,  \bar{q}}{1 - \bar{q}} \right)  
     + \frac{k_m}{2} \frac{\bar{q}}{1-(1-k_m)\bar{q}}  = 0
\end{aligned}
\right.
\end{equation}
that are exactly Eqs.~\eqref{eq:sp1} with $k_m$ instead of $\bar{m}$ and $\bar{q}$ instead of $\bar{q}_1$. Therefore $k_m = \bar{m}$ and $\bar{q} = \bar{q}_1$ and one can check that
\begin{equation}
    \frac{g(k; 0, \bar{q}, k_m)}{k} = \frac{g_{0} (k_m, q)}{k_m} \,.
\end{equation}
It only remains to prove that $k_c$ and $k_m$, which in general can be different points, are actually the same. As the 1RSB ansatz gives the correct solution for the present model, the corresponding SCGF must be convex and thus, in particular, continuous. The only way to obtain a continuous function which is equal to the RS one above $k_c$ and to the Rammal's constant below, is to take $k_c = k_m$, and so the two functions coincide everywhere.

\bibliography{biblioLD}

\begin{thebibliography}{29}%
\makeatletter
\providecommand \@ifxundefined [1]{%
 \@ifx{#1\undefined}
}%
\providecommand \@ifnum [1]{%
 \ifnum #1\expandafter \@firstoftwo
 \else \expandafter \@secondoftwo
 \fi
}%
\providecommand \@ifx [1]{%
 \ifx #1\expandafter \@firstoftwo
 \else \expandafter \@secondoftwo
 \fi
}%
\providecommand \natexlab [1]{#1}%
\providecommand \enquote  [1]{``#1''}%
\providecommand \bibnamefont  [1]{#1}%
\providecommand \bibfnamefont [1]{#1}%
\providecommand \citenamefont [1]{#1}%
\providecommand \href@noop [0]{\@secondoftwo}%
\providecommand \href [0]{\begingroup \@sanitize@url \@href}%
\providecommand \@href[1]{\@@startlink{#1}\@@href}%
\providecommand \@@href[1]{\endgroup#1\@@endlink}%
\providecommand \@sanitize@url [0]{\catcode `\\12\catcode `\$12\catcode
  `\&12\catcode `\#12\catcode `\^12\catcode `\_12\catcode `\%12\relax}%
\providecommand \@@startlink[1]{}%
\providecommand \@@endlink[0]{}%
\providecommand \url  [0]{\begingroup\@sanitize@url \@url }%
\providecommand \@url [1]{\endgroup\@href {#1}{\urlprefix }}%
\providecommand \urlprefix  [0]{URL }%
\providecommand \Eprint [0]{\href }%
\providecommand \doibase [0]{http://dx.doi.org/}%
\providecommand \selectlanguage [0]{\@gobble}%
\providecommand \bibinfo  [0]{\@secondoftwo}%
\providecommand \bibfield  [0]{\@secondoftwo}%
\providecommand \translation [1]{[#1]}%
\providecommand \BibitemOpen [0]{}%
\providecommand \bibitemStop [0]{}%
\providecommand \bibitemNoStop [0]{.\EOS\space}%
\providecommand \EOS [0]{\spacefactor3000\relax}%
\providecommand \BibitemShut  [1]{\csname bibitem#1\endcsname}%
\let\auto@bib@innerbib\@empty
\bibitem [{\citenamefont {{Toulouse, G.}}\ and\ \citenamefont {{Derrida,
  B.}}(1981)}]{TD_old}%
  \BibitemOpen
  \bibfield  {author} {\bibinfo {author} {\bibnamefont {{Toulouse, G.}}}\ and\
  \bibinfo {author} {\bibnamefont {{Derrida, B.}}},\ }in\ \href
  {http://inis.iaea.org/search/search.aspx?orig_q=RN:14717630} {\emph {\bibinfo
  {booktitle} {Proceedings of the Sixth Symposium on Theoretical Physics}}}\
  (\bibinfo {address} {Rio de Janeiro, Brazil},\ \bibinfo {year} {1981})\ p.\
  \bibinfo {pages} {217}\BibitemShut {NoStop}%
\bibitem [{\citenamefont {Tanaka}\ \emph {et~al.}(1989)\citenamefont {Tanaka},
  \citenamefont {Fujisaka},\ and\ \citenamefont {Inoue}}]{Tanaka}%
  \BibitemOpen
  \bibfield  {author} {\bibinfo {author} {\bibfnamefont {T.}~\bibnamefont
  {Tanaka}}, \bibinfo {author} {\bibfnamefont {H.}~\bibnamefont {Fujisaka}}, \
  and\ \bibinfo {author} {\bibfnamefont {M.}~\bibnamefont {Inoue}},\ }\href
  {\doibase 10.1103/PhysRevA.39.3170} {\bibfield  {journal} {\bibinfo
  {journal} {Phys. Rev. A}\ }\textbf {\bibinfo {volume} {39}},\ \bibinfo
  {pages} {3170} (\bibinfo {year} {1989})}\BibitemShut {NoStop}%
\bibitem [{\citenamefont {Crisanti}\ \emph {et~al.}(1990)\citenamefont
  {Crisanti}, \citenamefont {Nicolis}, \citenamefont {Paladin},\ and\
  \citenamefont {Vulpiani}}]{Crisanti_1990}%
  \BibitemOpen
  \bibfield  {author} {\bibinfo {author} {\bibfnamefont {A.}~\bibnamefont
  {Crisanti}}, \bibinfo {author} {\bibfnamefont {S.}~\bibnamefont {Nicolis}},
  \bibinfo {author} {\bibfnamefont {G.}~\bibnamefont {Paladin}}, \ and\
  \bibinfo {author} {\bibfnamefont {A.}~\bibnamefont {Vulpiani}},\ }\href
  {\doibase 10.1088/0305-4470/23/13/042} {\bibfield  {journal} {\bibinfo
  {journal} {Journal of Physics A: Mathematical and General}\ }\textbf
  {\bibinfo {volume} {23}},\ \bibinfo {pages} {3083} (\bibinfo {year}
  {1990})}\BibitemShut {NoStop}%
\bibitem [{\citenamefont {Rivoire}(2005)}]{RivoireCAV}%
  \BibitemOpen
  \bibfield  {author} {\bibinfo {author} {\bibfnamefont {O.}~\bibnamefont
  {Rivoire}},\ }\href
  {https://iopscience.iop.org/article/10.1088/1742-5468/2005/07/P07004/fulltext/}
  {\bibfield  {journal} {\bibinfo  {journal} {Journal of Statistical Mechanics:
  Theory and Experiment}\ }\textbf {\bibinfo {volume} {2005}},\ \bibinfo
  {pages} {P07004} (\bibinfo {year} {2005})}\BibitemShut {NoStop}%
\bibitem [{\citenamefont {Parisi}\ and\ \citenamefont
  {Rizzo}(2008)}]{rizzoPRL}%
  \BibitemOpen
  \bibfield  {author} {\bibinfo {author} {\bibfnamefont {G.}~\bibnamefont
  {Parisi}}\ and\ \bibinfo {author} {\bibfnamefont {T.}~\bibnamefont {Rizzo}},\
  }\href {\doibase 10.1103/PhysRevLett.101.117205} {\bibfield  {journal}
  {\bibinfo  {journal} {Phys. Rev. Lett.}\ }\textbf {\bibinfo {volume} {101}},\
  \bibinfo {pages} {117205} (\bibinfo {year} {2008})}\BibitemShut {NoStop}%
\bibitem [{\citenamefont {Parisi}\ and\ \citenamefont
  {Rizzo}(2009)}]{rizzoPRB2008}%
  \BibitemOpen
  \bibfield  {author} {\bibinfo {author} {\bibfnamefont {G.}~\bibnamefont
  {Parisi}}\ and\ \bibinfo {author} {\bibfnamefont {T.}~\bibnamefont {Rizzo}},\
  }\href {\doibase 10.1103/PhysRevB.79.134205} {\bibfield  {journal} {\bibinfo
  {journal} {Phys. Rev. B}\ }\textbf {\bibinfo {volume} {79}},\ \bibinfo
  {pages} {134205} (\bibinfo {year} {2009})}\BibitemShut {NoStop}%
\bibitem [{\citenamefont {Parisi}\ and\ \citenamefont
  {Rizzo}(2010{\natexlab{a}})}]{rizzoPRB2010}%
  \BibitemOpen
  \bibfield  {author} {\bibinfo {author} {\bibfnamefont {G.}~\bibnamefont
  {Parisi}}\ and\ \bibinfo {author} {\bibfnamefont {T.}~\bibnamefont {Rizzo}},\
  }\href {\doibase 10.1103/PhysRevB.81.094201} {\bibfield  {journal} {\bibinfo
  {journal} {Phys. Rev. B}\ }\textbf {\bibinfo {volume} {81}},\ \bibinfo
  {pages} {094201} (\bibinfo {year} {2010}{\natexlab{a}})}\BibitemShut
  {NoStop}%
\bibitem [{\citenamefont {Parisi}\ and\ \citenamefont
  {Rizzo}(2010{\natexlab{b}})}]{rizzoJPA}%
  \BibitemOpen
  \bibfield  {author} {\bibinfo {author} {\bibfnamefont {G.}~\bibnamefont
  {Parisi}}\ and\ \bibinfo {author} {\bibfnamefont {T.}~\bibnamefont {Rizzo}},\
  }\href {http://stacks.iop.org/1751-8121/43/i=4/a=045001} {\bibfield
  {journal} {\bibinfo  {journal} {Journal of Physics A: Mathematical and
  Theoretical}\ }\textbf {\bibinfo {volume} {43}},\ \bibinfo {pages} {045001}
  (\bibinfo {year} {2010}{\natexlab{b}})}\BibitemShut {NoStop}%
\bibitem [{\citenamefont {Andreanov}\ \emph {et~al.}(2004)\citenamefont
  {Andreanov}, \citenamefont {Barbieri},\ and\ \citenamefont
  {Martin}}]{AndREM}%
  \BibitemOpen
  \bibfield  {author} {\bibinfo {author} {\bibfnamefont {A.}~\bibnamefont
  {Andreanov}}, \bibinfo {author} {\bibfnamefont {F.}~\bibnamefont {Barbieri}},
  \ and\ \bibinfo {author} {\bibfnamefont {O.}~\bibnamefont {Martin}},\ }\href
  {https://link.springer.com/article/10.1140/epjb/e2004-00329-0} {\bibfield
  {journal} {\bibinfo  {journal} {The European Physical Journal B-Condensed
  Matter and Complex Systems}\ }\textbf {\bibinfo {volume} {41}},\ \bibinfo
  {pages} {365} (\bibinfo {year} {2004})}\BibitemShut {NoStop}%
\bibitem [{\citenamefont {Nakajima}\ and\ \citenamefont
  {Hukushima}(2008)}]{nakhukSK}%
  \BibitemOpen
  \bibfield  {author} {\bibinfo {author} {\bibfnamefont {T.}~\bibnamefont
  {Nakajima}}\ and\ \bibinfo {author} {\bibfnamefont {K.}~\bibnamefont
  {Hukushima}},\ }\href {\doibase 10.1143/JPSJ.77.074718} {\bibfield  {journal}
  {\bibinfo  {journal} {Journal of the Physical Society of Japan}\ }\textbf
  {\bibinfo {volume} {77}},\ \bibinfo {pages} {074718} (\bibinfo {year}
  {2008})},\ \Eprint
  {http://arxiv.org/abs/https://doi.org/10.1143/JPSJ.77.074718}
  {https://doi.org/10.1143/JPSJ.77.074718} \BibitemShut {NoStop}%
\bibitem [{\citenamefont {Nakajima}\ and\ \citenamefont
  {Hukushima}(2009)}]{nakhukPRE}%
  \BibitemOpen
  \bibfield  {author} {\bibinfo {author} {\bibfnamefont {T.}~\bibnamefont
  {Nakajima}}\ and\ \bibinfo {author} {\bibfnamefont {K.}~\bibnamefont
  {Hukushima}},\ }\href {\doibase 10.1103/PhysRevE.80.011103} {\bibfield
  {journal} {\bibinfo  {journal} {Phys. Rev. E}\ }\textbf {\bibinfo {volume}
  {80}},\ \bibinfo {pages} {011103} (\bibinfo {year} {2009})}\BibitemShut
  {NoStop}%
\bibitem [{\citenamefont {Malatesta}\ \emph {et~al.}(2019)\citenamefont
  {Malatesta}, \citenamefont {Parisi},\ and\ \citenamefont
  {Sicuro}}]{MalatestaLDMatch}%
  \BibitemOpen
  \bibfield  {author} {\bibinfo {author} {\bibfnamefont {E.~M.}\ \bibnamefont
  {Malatesta}}, \bibinfo {author} {\bibfnamefont {G.}~\bibnamefont {Parisi}}, \
  and\ \bibinfo {author} {\bibfnamefont {G.}~\bibnamefont {Sicuro}},\ }\href
  {\doibase 10.1103/PhysRevE.100.032102} {\bibfield  {journal} {\bibinfo
  {journal} {Phys. Rev. E}\ }\textbf {\bibinfo {volume} {100}},\ \bibinfo
  {pages} {032102} (\bibinfo {year} {2019})}\BibitemShut {NoStop}%
\bibitem [{\citenamefont {Touchette}(2009)}]{touchLD}%
  \BibitemOpen
  \bibfield  {author} {\bibinfo {author} {\bibfnamefont {H.}~\bibnamefont
  {Touchette}},\ }\href {\doibase
  https://doi.org/10.1016/j.physrep.2009.05.002} {\bibfield  {journal}
  {\bibinfo  {journal} {Physics Reports}\ }\textbf {\bibinfo {volume} {478}},\
  \bibinfo {pages} {1 } (\bibinfo {year} {2009})}\BibitemShut {NoStop}%
\bibitem [{\citenamefont {Pérez-Espigares}\ and\ \citenamefont
  {Hurtado}(2019)}]{Carlos:Chaos:2019}%
  \BibitemOpen
  \bibfield  {author} {\bibinfo {author} {\bibfnamefont {C.}~\bibnamefont
  {Pérez-Espigares}}\ and\ \bibinfo {author} {\bibfnamefont {P.~I.}\
  \bibnamefont {Hurtado}},\ }\href {\doibase 10.1063/1.5091669} {\bibfield
  {journal} {\bibinfo  {journal} {Chaos: An Interdisciplinary Journal of
  Nonlinear Science}\ }\textbf {\bibinfo {volume} {29}},\ \bibinfo {pages}
  {083106} (\bibinfo {year} {2019})},\ \Eprint
  {http://arxiv.org/abs/https://doi.org/10.1063/1.5091669}
  {https://doi.org/10.1063/1.5091669} \BibitemShut {NoStop}%
\bibitem [{\citenamefont {Parisi}(1980)}]{parisiRSB}%
  \BibitemOpen
  \bibfield  {author} {\bibinfo {author} {\bibfnamefont {G.}~\bibnamefont
  {Parisi}},\ }\href {http://stacks.iop.org/0305-4470/13/i=4/a=009} {\bibfield
  {journal} {\bibinfo  {journal} {Journal of Physics A: Mathematical and
  General}\ }\textbf {\bibinfo {volume} {13}},\ \bibinfo {pages} {L115}
  (\bibinfo {year} {1980})}\BibitemShut {NoStop}%
\bibitem [{\citenamefont {van Hemmen}\ and\ \citenamefont
  {Palmer}(1979)}]{vanH-P}%
  \BibitemOpen
  \bibfield  {author} {\bibinfo {author} {\bibfnamefont {J.~L.}\ \bibnamefont
  {van Hemmen}}\ and\ \bibinfo {author} {\bibfnamefont {R.~G.}\ \bibnamefont
  {Palmer}},\ }\href {http://stacks.iop.org/0305-4470/12/i=4/a=016} {\bibfield
  {journal} {\bibinfo  {journal} {Journal of Physics A: Mathematical and
  General}\ }\textbf {\bibinfo {volume} {12}},\ \bibinfo {pages} {563}
  (\bibinfo {year} {1979})}\BibitemShut {NoStop}%
\bibitem [{\citenamefont {Rammal}(1981)}]{Rammal}%
  \BibitemOpen
  \bibfield  {author} {\bibinfo {author} {\bibfnamefont {R.}~\bibnamefont
  {Rammal}},\ }\href@noop {} {\bibfield  {journal} {\bibinfo  {journal} {PhD
  thesis (unpublished), Grenoble University}\ } (\bibinfo {year}
  {1981})}\BibitemShut {NoStop}%
\bibitem [{\citenamefont {Kondor}(1983)}]{kond}%
  \BibitemOpen
  \bibfield  {author} {\bibinfo {author} {\bibfnamefont {I.}~\bibnamefont
  {Kondor}},\ }\href {http://stacks.iop.org/0305-4470/16/i=4/a=006} {\bibfield
  {journal} {\bibinfo  {journal} {Journal of Physics A: Mathematical and
  General}\ }\textbf {\bibinfo {volume} {16}},\ \bibinfo {pages} {L127}
  (\bibinfo {year} {1983})}\BibitemShut {NoStop}%
\bibitem [{\citenamefont {Gardner}\ and\ \citenamefont
  {Derrida}(1989)}]{gardnerREM}%
  \BibitemOpen
  \bibfield  {author} {\bibinfo {author} {\bibfnamefont {E.}~\bibnamefont
  {Gardner}}\ and\ \bibinfo {author} {\bibfnamefont {B.}~\bibnamefont
  {Derrida}},\ }\href {http://stacks.iop.org/0305-4470/22/i=12/a=003}
  {\bibfield  {journal} {\bibinfo  {journal} {Journal of Physics A:
  Mathematical and General}\ }\textbf {\bibinfo {volume} {22}},\ \bibinfo
  {pages} {1975} (\bibinfo {year} {1989})}\BibitemShut {NoStop}%
\bibitem [{\citenamefont {Fedrigo}\ \emph {et~al.}(2007)\citenamefont
  {Fedrigo}, \citenamefont {Flandoli},\ and\ \citenamefont
  {Morandin}}]{fedrigo2007}%
  \BibitemOpen
  \bibfield  {author} {\bibinfo {author} {\bibfnamefont {M.}~\bibnamefont
  {Fedrigo}}, \bibinfo {author} {\bibfnamefont {F.}~\bibnamefont {Flandoli}}, \
  and\ \bibinfo {author} {\bibfnamefont {F.}~\bibnamefont {Morandin}},\ }\href
  {\doibase 10.1007/s10231-006-0011-4} {\bibfield  {journal} {\bibinfo
  {journal} {Annali di Matematica Pura ed Applicata}\ }\textbf {\bibinfo
  {volume} {186}},\ \bibinfo {pages} {381} (\bibinfo {year}
  {2007})}\BibitemShut {NoStop}%
\bibitem [{\citenamefont {Ogure}\ and\ \citenamefont
  {Kabashima}(2004)}]{ogkabPTP}%
  \BibitemOpen
  \bibfield  {author} {\bibinfo {author} {\bibfnamefont {K.}~\bibnamefont
  {Ogure}}\ and\ \bibinfo {author} {\bibfnamefont {Y.}~\bibnamefont
  {Kabashima}},\ }\href {\doibase 10.1143/PTP.111.661} {\bibfield  {journal}
  {\bibinfo  {journal} {Progress of Theoretical Physics}\ }\textbf {\bibinfo
  {volume} {111}},\ \bibinfo {pages} {661} (\bibinfo {year}
  {2004})}\BibitemShut {NoStop}%
\bibitem [{\citenamefont {Ogure}\ and\ \citenamefont
  {Kabashima}(2009{\natexlab{a}})}]{ogkabJSM1}%
  \BibitemOpen
  \bibfield  {author} {\bibinfo {author} {\bibfnamefont {K.}~\bibnamefont
  {Ogure}}\ and\ \bibinfo {author} {\bibfnamefont {Y.}~\bibnamefont
  {Kabashima}},\ }\href {http://stacks.iop.org/1742-5468/2009/i=03/a=P03010}
  {\bibfield  {journal} {\bibinfo  {journal} {Journal of Statistical Mechanics:
  Theory and Experiment}\ }\textbf {\bibinfo {volume} {2009}},\ \bibinfo
  {pages} {P03010} (\bibinfo {year} {2009}{\natexlab{a}})}\BibitemShut
  {NoStop}%
\bibitem [{\citenamefont {Ogure}\ and\ \citenamefont
  {Kabashima}(2009{\natexlab{b}})}]{ogkabJSM2}%
  \BibitemOpen
  \bibfield  {author} {\bibinfo {author} {\bibfnamefont {K.}~\bibnamefont
  {Ogure}}\ and\ \bibinfo {author} {\bibfnamefont {Y.}~\bibnamefont
  {Kabashima}},\ }\href {http://stacks.iop.org/1742-5468/2009/i=05/a=P05011}
  {\bibfield  {journal} {\bibinfo  {journal} {Journal of Statistical Mechanics:
  Theory and Experiment}\ }\textbf {\bibinfo {volume} {2009}},\ \bibinfo
  {pages} {P05011} (\bibinfo {year} {2009}{\natexlab{b}})}\BibitemShut
  {NoStop}%
\bibitem [{\citenamefont {Crisanti}\ and\ \citenamefont {Sommers}(1992)}]{CS}%
  \BibitemOpen
  \bibfield  {author} {\bibinfo {author} {\bibfnamefont {A.}~\bibnamefont
  {Crisanti}}\ and\ \bibinfo {author} {\bibfnamefont {H.~J.}\ \bibnamefont
  {Sommers}},\ }\href {\doibase 10.1007/BF01309287} {\bibfield  {journal}
  {\bibinfo  {journal} {Zeitschrift f{\"u}r Physik B Condensed Matter}\
  }\textbf {\bibinfo {volume} {87}},\ \bibinfo {pages} {341} (\bibinfo {year}
  {1992})}\BibitemShut {NoStop}%
\bibitem [{\citenamefont {Dotsenko}\ \emph {et~al.}(1994)\citenamefont
  {Dotsenko}, \citenamefont {Franz},\ and\ \citenamefont
  {Mezard}}]{Dotsenko_1994}%
  \BibitemOpen
  \bibfield  {author} {\bibinfo {author} {\bibfnamefont {V.}~\bibnamefont
  {Dotsenko}}, \bibinfo {author} {\bibfnamefont {S.}~\bibnamefont {Franz}}, \
  and\ \bibinfo {author} {\bibfnamefont {M.}~\bibnamefont {Mezard}},\ }\href
  {\doibase 10.1088/0305-4470/27/7/016} {\bibfield  {journal} {\bibinfo
  {journal} {Journal of Physics A: Mathematical and General}\ }\textbf
  {\bibinfo {volume} {27}},\ \bibinfo {pages} {2351} (\bibinfo {year}
  {1994})}\BibitemShut {NoStop}%
\bibitem [{\citenamefont {Kosterlitz}\ \emph {et~al.}(1976)\citenamefont
  {Kosterlitz}, \citenamefont {Thouless},\ and\ \citenamefont
  {Jones}}]{PhysRevLett.36.1217}%
  \BibitemOpen
  \bibfield  {author} {\bibinfo {author} {\bibfnamefont {J.~M.}\ \bibnamefont
  {Kosterlitz}}, \bibinfo {author} {\bibfnamefont {D.~J.}\ \bibnamefont
  {Thouless}}, \ and\ \bibinfo {author} {\bibfnamefont {R.~C.}\ \bibnamefont
  {Jones}},\ }\href {\doibase 10.1103/PhysRevLett.36.1217} {\bibfield
  {journal} {\bibinfo  {journal} {Phys. Rev. Lett.}\ }\textbf {\bibinfo
  {volume} {36}},\ \bibinfo {pages} {1217} (\bibinfo {year}
  {1976})}\BibitemShut {NoStop}%
\bibitem [{\citenamefont {Dean}\ and\ \citenamefont {Majumdar}(2008)}]{DeanRM}%
  \BibitemOpen
  \bibfield  {author} {\bibinfo {author} {\bibfnamefont {D.~S.}\ \bibnamefont
  {Dean}}\ and\ \bibinfo {author} {\bibfnamefont {S.~N.}\ \bibnamefont
  {Majumdar}},\ }\href
  {https://journals.aps.org/pre/abstract/10.1103/PhysRevE.77.041108} {\bibfield
   {journal} {\bibinfo  {journal} {Physical Review E}\ }\textbf {\bibinfo
  {volume} {77}},\ \bibinfo {pages} {041108} (\bibinfo {year}
  {2008})}\BibitemShut {NoStop}%
\bibitem [{\citenamefont {Dotsenko}(2011)}]{Dotsenko_2011}%
  \BibitemOpen
  \bibfield  {author} {\bibinfo {author} {\bibfnamefont {V.}~\bibnamefont
  {Dotsenko}},\ }\href {\doibase 10.1209/0295-5075/95/50006} {\bibfield
  {journal} {\bibinfo  {journal} {{EPL} (Europhysics Letters)}\ }\textbf
  {\bibinfo {volume} {95}},\ \bibinfo {pages} {50006} (\bibinfo {year}
  {2011})}\BibitemShut {NoStop}%
\bibitem [{\citenamefont {de~Almeida}\ and\ \citenamefont
  {Thouless}(1978)}]{dAT}%
  \BibitemOpen
  \bibfield  {author} {\bibinfo {author} {\bibfnamefont {J.~R.~L.}\
  \bibnamefont {de~Almeida}}\ and\ \bibinfo {author} {\bibfnamefont {D.~J.}\
  \bibnamefont {Thouless}},\ }\href
  {http://stacks.iop.org/0305-4470/11/i=5/a=028} {\bibfield  {journal}
  {\bibinfo  {journal} {Journal of Physics A: Mathematical and General}\
  }\textbf {\bibinfo {volume} {11}},\ \bibinfo {pages} {983} (\bibinfo {year}
  {1978})}\BibitemShut {NoStop}%
\end{thebibliography}%

\end{document}